\documentclass[11pt,preprint]{aastex}

\usepackage{amsmath,amstext}
\usepackage[T1]{fontenc}
\usepackage{graphicx}
\usepackage{amsmath,amssymb}
\usepackage{color}
\usepackage{bm}
\usepackage[breaklinks,colorlinks,citecolor=blue,linkcolor=blue]{hyperref} 
\usepackage[all]{hypcap} 
\usepackage[normalem]{ulem}
\usepackage{soul}

\bibpunct{(}{)}{;}{a}{}{,}

\def\mbh{M_{\bullet}}
\def\msun{M_{\odot}}
\def\kms{\rm km\,s^{-1}}
\def\cmc{\rm cm^{-3}}
\def\ergs{\rm erg\,s^{-1}}

\def\kb{k_{_{\rm B}}}
\def\rs{R_{\rm s}}
\def\vs{v_{\rm s}}

\def\dms{\dot{M}_{\rm s}}
\def\tdyn{t_{\rm dyn}}
\def\tff{t_{\rm ff}}
\def\tcool{t_{\rm cool}}

\def\Latm{\Lambda_{\rm atm}}
\def\ncl{n_{\rm cl}}
\def\tcl{T_{\rm cl}}

\def\rhog{\rho_{\rm g}}

\def\fs{f_{\star}}
\def\dmstar{\dot{M}_{\star}}
\def\mj{M_{\rm J}}

\def\taums{\tau_{\rm ms}}

\def\tsal{t_{\rm Sal}}
\def\mpr{m_{\rm p}}

\def\nc{n_{\rm c}}
\def\tc{T_{\rm c}}
\def\tatm{t_{\rm atm}}
\def\lti{l_{_{\rm TI}}}
\def\mti{M_{_{\rm TI}}}
\def\lcl{l_{\rm cl}}
\def\ltid{l_{\rm tid}}
\def\mcl{M_{\rm cl}}
\def\tpl{T_{\rm pl}}
\def\npl{n_{\rm pl}}
\def\rcl{R_{\rm cl}}
\def\ncl{n_{\rm cl}}
\def\tcl{T_{\rm cl}}
\def\cs{c_{\rm s}}
\def\meanms{\langle M_{\star}\rangle}
\def\mmin{M_{\star,\rm min}}
\def\mmax{M_{\star,\rm max}}
\def\sigstar{\sigma_{\star}}

\def\th{T_{\rm h}}
\def\nh{n_{\rm h}}

\title{Formation and spatial distribution of hypervelocity stars in AGN outflows}
\author{Xiawei Wang and Abraham Loeb}
\affil{Department of Astronomy, Harvard University, 60 Garden Street, Cambridge, MA 02138, USA}%

\begin{abstract}
We study star formation within outflows driven by active galactic nuclei (AGN) as a new source of hypervelocity stars (HVSs).
Recent observations revealed active star formation inside a galactic outflow at a rate of $\sim 15\,\msun\,\rm yr^{-1}$.
We verify that the shells swept up by an AGN outflow are capable of cooling and fragmentation into cold clumps embedded in a hot tenuous gas via thermal instabilities.
We show that cold clumps of $\sim 10^3\,\msun$ are formed within $\sim 10^5$ yrs.
As a result, stars are produced along outflow's path, endowed with the outflow speed at their formation site.
These HVSs travel through the galactic halo and eventually escape into the intergalactic medium.
The expected instantaneous rate of star formation inside the outflow is $\sim 4-5$ orders of magnitude greater than the average rate associated with previously proposed mechanisms for producing HVSs, such as the Hills mechanism and three-body interaction between a star and a black hole binary.
We predict the spatial distribution of HVSs formed in AGN outflows for future observational probe.
\end{abstract}
\keywords
{
galaxies: active --
stars: formation --
ISM: jets and outflows
}

\begin{document}
%
%
\section{Introduction} 
\label{sec:intro}
There is growing observational evidence for large scale outflows driven by active galactic nuclei (AGNs).
Such outflows have been detected in nearby ultra luminous infrared galaxies \citep{cicone2014, tombesi2015} as well as in broad absorption line quasars \citep{arav2015}.
Interestingly, cold molecular clumps are observed through their CO and HCN emission to co-exist with hot gas in outflows, forming a multi-phase medium in equilibrium \citep{cicone2014}.
Recent observations of a nearby galaxy revealed possible ongoing star formation inside a massive galactic outflow for the first time \citep{maiolino2017}.

Previously, \citet{silk2012} discussed the ejection of hypervelocity stars (HVSs) in the Galactic Center as a result of AGN jet interaction with a giant molecular cloud.
In addition, it has been discussed that AGN outflows can trigger or enhance star formation by compressing pre-existing cold gas in the interstellar medium (ISM), such as \citet{ishibashi2012, ishibashi2014, nayakshin2012, silk2013, zubovas2013, zubovas2017}.
Instead, we focus here on a different scenario of HVS production where the outflow material itself fragments into stars late in the hydrodynamical evolution of the outflow.
Numerical simulations have identified the required physical conditions for the formation of molecular clumps in AGN outflows \citep{costa2015, ferrara2016, scannapieco2017, richings2017}, due to a thermal instability. 
The resulting distribution of stars could be substantially different from the previously considered scenario since stars are born with the outflow's speed in this case.
Here we calculate cold clump formation in detail, and discuss the detailed properties and statistics of the resulting stellar population and spatial distribution, which has not been considered in the literature (e.g. \citet{zubovas2014, zubovas2017}).

Over the past decade, dozens of HVSs have been detected in the halo of the Milky Way (MW) galaxy \citep{brown2015}.
The fastest known stars have velocities $\sim 700\,\kms$ at distances of $50-100$ kpc \citep{brown2014}, which significantly exceed the escape speed of the MW halo. 
Unbound HVSs are distributed equally across Galactic latitude but appear clumped in Galactic longitude \citep{brown2009, boubert2016, boubert2017}.
The spatial and velocity distribution of identified HVSs suggest a scenario of three-body exchange in which the supermassive black hole at the Galactic center (GC), Sgr A*, dissociates through its gravitational tide a binary star system and ejects one of its members as a HVS, in a process known as the Hills mechanism \citep{hills1988, brown2015}.

Here we verify that star formation in AGN outflows could lead to an alternative production channel of HVSs, at a rate of $\sim10\,\msun\,\rm yr^{-1}$, consistent with the observed rate \citep{maiolino2017}, which is $\sim 4-5$ orders of magnitude greater than the Hills mechanism \citep{hills1988}, as well as other previously considered processes, such as three-body interaction between a star and a binary black hole system \citep{yu2003, guillochon2015}.
We discuss formation of cold clumps via thermal instabilities in detail, and predict the spatial distribution of HVSs formed in AGN outflows for future observational probe, which has not been discussed in previous literature (e.g. \citet{zubovas2014}).

Our paper is organized as follows.
In \S\ref{sec:two-phase}, we discuss the AGN outflow hydrodynamics and the formation of cold clumps.
In \S\ref{sec:HVS}, we discuss star formation within the outflow and predict the statistics of hypervelocity stars born in an AGN outflow.
Finally, in \S\ref{sec:summary} we summarize our results and discuss related observational implications.
%
%
\section{Two-phase medium}
\label{sec:two-phase}
\subsection{Outflow hydrodynamics}
\label{subsec:hydro}
\begin{figure}[h!]
\centering
\includegraphics[angle=0,width=1\columnwidth]{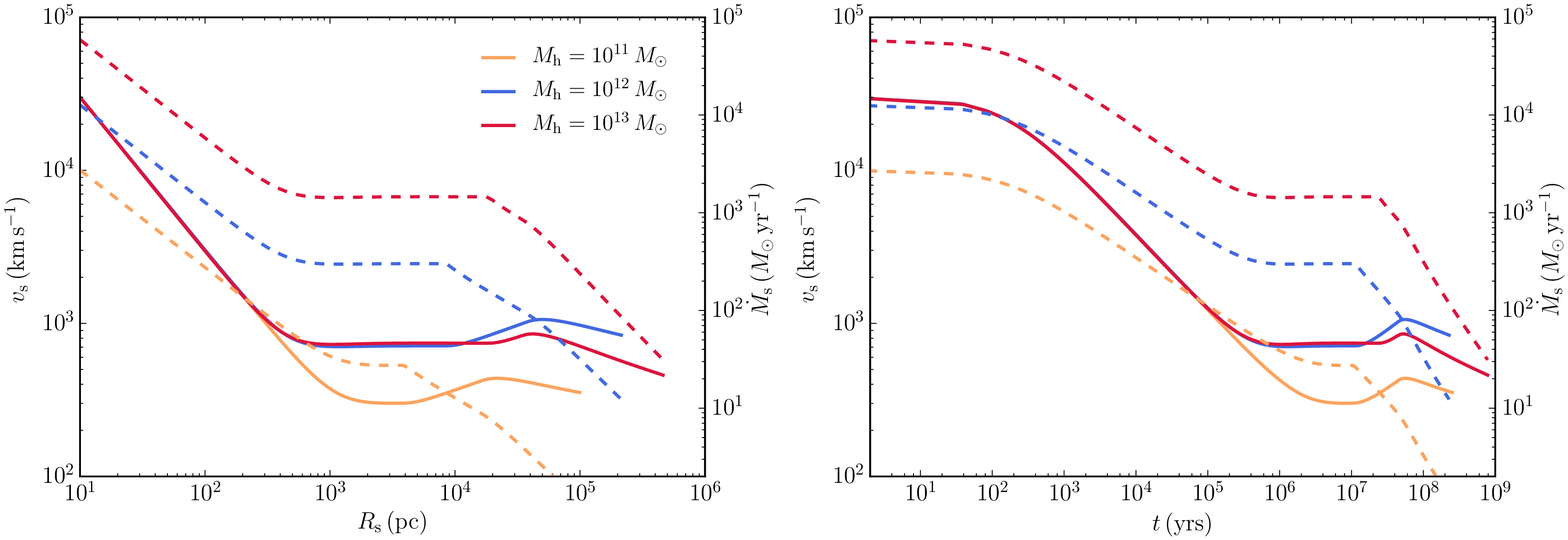}
\caption{
Hydrodynamics of AGN outflows embedded in halos of $M_{\rm h}=10^{11},10^{12},10^{13}\,\msun$ (represented by orange, blue and red lines, respectively).
In the left and right panels, we show the outflow speed, $\vs$ (solid lines) and the interception rate of swept-up mass, $\dms$ (dashed lines), as functions of the outflow radius, $\rs$ and time, $t$.
The black hole mass, $\mbh$, and the gas density distribution are self-consistently determined by $M_{\rm h}$ and redshift $z$ (see \citet{wang2015} for details).
For the outflow parameters of interest, we find that the outflow can reach the edge of the halo with a speed of $\gtrsim 300\,\kms$ on a timescale of $\sim 10^8$ yrs.
}
\label{fig:fig1}
\end{figure}

AGNs are believed to launch a fast wind from their inner accretion disk with a velocity of $\sim 0.1\,c$, where $c$ is the speed of light \citep{king2015}. 
The wind drives a double shock structure, where the outer forward shock sweeps up the ambient medium while the inner reverse shock decelerates the wind itself.
The two shocks are separated at a contact discontinuity.
Here we follow the hydrodynamical model of outflow's outer boundary from our previous work (see \citet{wang2015} for details).
The continuous energy injection into the wind is assumed to last for a Salpeter time, $\tsal\sim 4.5\times 10^7$ yrs, after which the AGN shuts off, assuming a radiative efficiency from accretion of $\sim10\%$.
We adopt a broken power-law radial density profile for the gas, $\rhog$, which follows an isothermal sphere and NFW profiles in the disk and halo components, respectively. 
Figure \ref{fig:fig1} shows the hydrodynamics of the outflows embedded in halos of $10^{11}, 10^{12}, 10^{13}\,\msun$.
The speed of the outflowing shell, $\vs$, rapidly declines to a few hundreds of $\kms$ when it reaches the edge of the galactic disk and enters the halo.
The outflow continues to propagate into the halo even after the energy injection from the AGN shuts off.

\subsection{Clump formation}
\label{subsec:clump}
AGN outflows have been observed to be energy-conserving on large scales, where radiative cooling by the shocked wind is negligible \citep{tombesi2015}, a result supported by theoretical models \citep{fgq2012, king2015, wang2015}.
We note that it is the cooling of the shocked wind, not the shocked ambient medium, that determines whether the outflow is energy or momentum-conserving.
In analogy to supernova remnants, protons and electrons in the shocked wind region of AGN outflows can be significantly decoupled. 
The thermal energy carried by the protons is thus trapped in the shocked wind, leading to energy conservation.
The final temperature of the plasma in the shocked wind region reaches $\sim 10^7$ K, and remains too hot for clump condensation \citep{fgq2012}.
Previous simulations showed that during the dynamical transition between the momentum-driven and energy-driven phases, the shell is accelerated and fragmented by Rayleigh-Taylor instabilities, resulting in short-lived clumps entrained and ablated by hot gas flowing past them \citep{ferrara2016}.
Once the outflow becomes energy-conserving, the swept-up shell cools rapidly and condenses, with additional cold clumps formed via a thermal instability.
These clumps are able to survive as they are nearly at rest with the hot tenuous gas surrounding them \citep{scannapieco2017}.
Here we estimate the cooling timescale and final temperature of the clumps condensing out of the outflowing shell, and adopt prescriptions for the related heating and cooling functions \citep{sazonov2005, koyama2002}.

At $T\gtrsim 10^4$ K, the heating and cooling of the swept-up shell involves free-free emission, Compton heating/cooling, photoionization heating, line and recombination continuum cooling.
We adopt numerical approximations for relevant heating and cooling curves from \citet{sazonov2005} for optically thin gas illuminated by quasar radiation.
At $T\lesssim 10^4$ K, the dominant cooling process includes atomic and molecular cooling. 
We adopt related prescriptions from \citet{koyama2002} at solar metallicity:
\begin{equation}
\frac{\Lambda(T)}{\Gamma}=10^7\exp\left(-\frac{114800}{T+1000}\right) + 14\sqrt{T}\exp\left(-\frac{92}{T}\right)\;,
\end{equation}
where $\Gamma=2\times10^{-26}\,\ergs$ and $T$ is in units of K.
The above formula includes the following processes: photoelectric heating from small grains and polycyclic aromatic hydrocarbons, heating and ionization by cosmic rays and X-rays, heating by $\rm H_2$ formation and destruction, atomic line cooling from hydrogen $\rm Ly\alpha$, C $\textsc{ii}$, O $\textsc{i}$, Fe $\textsc{ii}$ and Si $\textsc{ii}$, rovibrational line cooling from $\rm H_2$ and CO, and atomic and molecular collisions with dust grains at solar metallicity.
The rate at which the energy density of the outflowing gas per unit volume changes due to the heating and cooling processes can be written as:
\begin{equation}
\dot{E}=H(n,T)-C(n,T)\;,
\end{equation}
where $H(n,T)$ and $C(n,T)$ are the total heating and cooling functions, respectively.
We numerically integrate the energy balance equation, starting from an initial temperature of the swept-up shell set by the Rankine-Hugoniot jump condition: $T_0\approx3\mu\mpr\vs^2/16\kb\approx10^7\,v_{\rm s,3}^2$ K, where $v_{\rm s,3}=(\vs/10^3\,\kms)$, $\mu=0.5$ is the mean molecular weight of fully ionized gas, $\mpr$ is the proton mass and $\kb$ is the Boltzmann constant.
\begin{figure}[h!]
\centering
\includegraphics[angle=0,width=0.9\columnwidth]{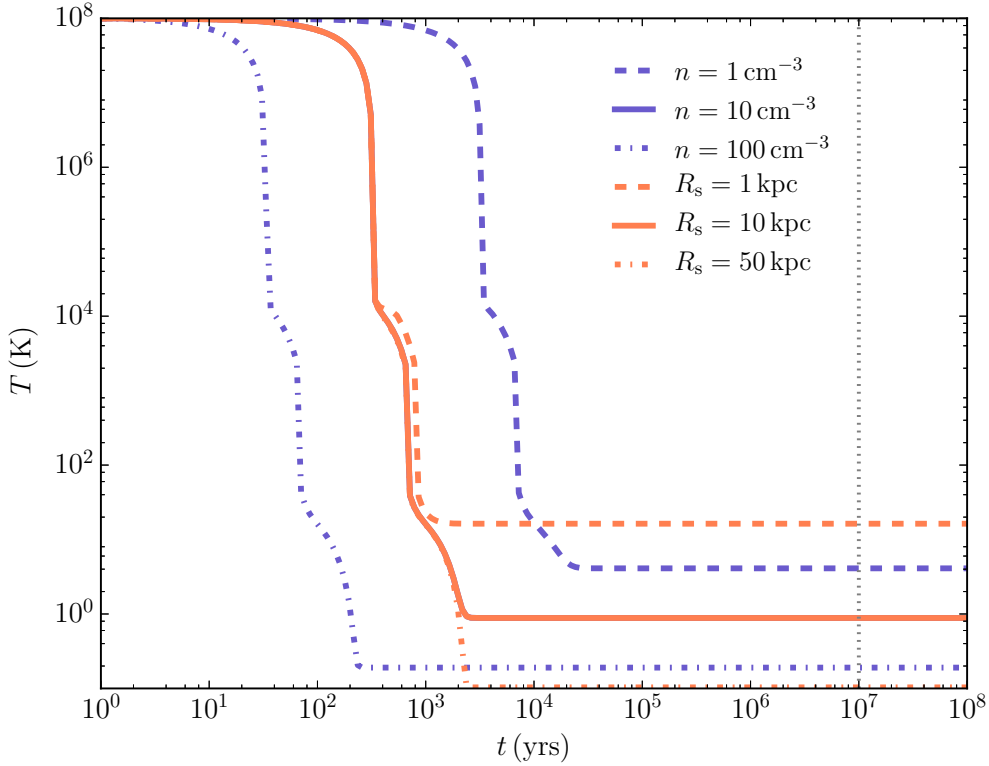}
\caption{
Temperature evolution of the swept-up gas shell.
We integrate the heating and cooling function that includes free-free cooling, Compton heating/cooling, photoionization heating, line and recombination continuum cooling, and atomic and molecular cooling.
The orange and purple lines show the temperature of the cooling gas as a function of time in a MW mass halo for different gas number density $n$ and different outflow radii $\rs$.
The AGN luminosity is fixed here to $\sim 10^{43}\,\ergs$.
The dotted grey line represents a characteristic outflow dynamical timescale, $\tdyn\sim 10^7$ yrs.
}
\label{fig:fig2}
\end{figure}

As shown in Fig.\ref{fig:fig2}, we find that for a range of parameter values of interest, the swept-up shell can cool down to $\sim 10$ K on a timescale of $\lesssim 10^4$ yrs, approximately scaled as $\tcool\sim10^4\,(n/\rm cm^{-3})^{-1}$ yrs, which is much shorter than the dynamical timescale of the outflow, $\tdyn\sim \rs/\vs\approx 10^7 R_{\rm s,1}\, v_{\rm s,3}\,\rm yrs$, where $R_{\rm s,1}=(\rs/10\,\rm kpc)$.
For comparison, the atomic and molecular cooling timescale is of order, $\tatm\sim 3\nc k \tc/\Latm\approx 1.3\times10^{-3}\,n_{\rm c,0}\,\rm yrs$,
where $n_{\rm c,0}=(\nc/1\,\cmc)$.
$\Latm\sim10^{-25}\,\nc^2\,\ergs\cmc$ is the cooling rate \citep{spitzer1978}.
Thus, the shocked ambient gas cools efficiently to temperatures amenable to clump condensation and subsequent star formation.
Thermal instability (TI) occurs if the heating rate at a constant pressure rises faster than the cooling rate \citep{field1965}.
Material slightly cooler than the surrounding medium will keep cooling down at a constant pressure if the following condition holds \citep{beltrametti1981}:
\begin{equation}
\left[\frac{\partial}{\partial T}(H-C)\right]_p >0\;.
\end{equation}
We have verified that for $T\sim 10^4$ K and $n\sim10^2\,\cmc$, the critical galactocentric distance above which the shell becomes thermally unstable is $r_{\rm c}\sim 1$ kpc.
For $T\sim 10^4$ K and $n\sim0.1\,\cmc$, $r_{\rm c}\sim 0.1$ kpc.
Cold clumps condense out of the outflowing shell on a timescale much shorter than outflow's dynamical timescale.
For a hot plasma with temperature of $\th$ and density $\nh$, the minimum size of the cold clumps is constrained by thermal conductivity to the value \citep{beltrametti1981}:
\begin{equation}
\lti=\left(2\kappa_0 \tpl/5c_0 \npl^2\right)^{1/2}=33.2\,q_0^{1/2}T_{\rm h,7}^{3/2}n_{\rm h,1}^{-1}\,\rm pc\;,
\end{equation}
where $T_{\rm h,7}=(\th/10^7\rm\, K)$, $n_{\rm h,1}=(\nh/10\,\rm cm^{-3})$, $\kappa_0=2\times10^{12}q_0 T_{\rm h,7}^{5/2}$ is the thermal conductivity, and $q_0=1+0.0015\ln(T_{\rm h,7}/n_{\rm h,1})$.
We adopt $q_0\approx1$ in the calculation.
The Bremsstrahlung coefficient is $c_0=7.6\times10^{-24}\,T_{\rm h,7}^{1/2}\,\rm ergs\,cm^{3}\,s^{-1}$.
For clumps of size $\lesssim\lti$, thermal conduction will efficiently transport heat to prevent the growth of TI.
The corresponding lower limit of gas mass enclosed in $\lcl$ can be written as:
\begin{equation}
\label{eq:eq9}
\mti=\left(\frac{4\pi}{3}\right)\lti^3\,\nh\mpr=3.5\times10^4\,T_{\rm h,7}^{9/2}n_{\rm h,1}^{-2}\,\msun
\end{equation}
During clump contraction, the size and density of the resulting clumps can be estimated from mass conservation $\ncl\rcl^3=\nh\lti^3$ and pressure balance $\ncl\tcl=\nh\th$, yielding:
\begin{equation}
\rcl\approx0.33\,T_{\rm cl,1}^{1/3} T_{\rm h,7}^{7/6}n_{\rm h,1}^{-1}\,\rm pc\;,
\end{equation}
where $T_{\rm cl,1}=(\tcl/10\rm K)$ is the temperature of the cold clumps, and
\begin{equation}
\ncl\approx10^7 T_{\rm cl,1} T_{\rm h,7}n_{\rm h,1}\,\rm cm^{-3}\;.
\end{equation}
The timescale of clump contraction driven by the surrounding medium can be obtained from:
\begin{equation}
t_{\rm shrink}\sim\lti/\cs\approx10^5 T_{\rm h,7} n_{\rm h,1}^{-1}\,\rm yrs\;,
\end{equation}
where $\cs$ is the adiabatic sound speed.
We find that $t_{\rm shrink}$ is significantly longer than the cooling timescale, $\tcool$, but much shorter than outflow's dynamical timescale, $\tdyn$.
Thus, the TI-formed clumps contract to reach their minimum size, determined by either thermal pressure or turbulent pressure.
An additional constraint on the size of the TI-formed clumps is associated with the tidal force from mass within outflow's radius, $\rs$.
The corresponding clump size can be expressed as:
\begin{equation}
\ltid=\left(\frac{\mti}{M_{\star}}\right)^{1/3}\,\rs=10.2\left(\frac{M_{\rm TI,4}}{M_{\star,10}}\right)^{1/3}\,R_{\rm s,kpc}\,\rm pc\;,
\end{equation}
where $M_{\rm TI,4}=(\mti/10^4\,\msun)$, $M_{\star,10}=(M_{\star}/10^{10}\,\msun)$ and $R_{\rm s,kpc}=(\rs/\rm kpc)$.
For $\mti\sim 3.5\times10^{4}\,\msun$, $\ltid\approx15.5\,\rm pc$$\sim\lti$, indicating that the clumps can survive the potential of galactic bulge.
We note that the magnetic field in the ISM could modify both the amplitude and morphology of thermal instability.
Recent numerical simulations show that magnetic tension suppresses buoyant oscillations of condensing gas, thus enhancing thermal instability \citep{ji2017}. 
The density fluctuation amplitude scales as $\delta\rho/\rho\propto\beta^{-1/2}$, where $\beta$ is the ratio between thermal and magnetic pressure, independent of the magnetic field orientation \citep{ji2017}.
Therefore, the scale constraints of the TI-induced clouds estimated here provide a lower limit on the growth of the clouds.
%
%
\section{Hypervelocity stars}
\label{sec:HVS}
\subsection{Star formation}
\label{subsec:star formation}
Clumps with a mass $\mcl\sim\mti\approx3.5\times10^4\,T_{\rm h,7}^{9/2}n_{\rm h,1}^{-2}\,\msun$ and size $\rcl\approx0.33 T_{\rm cl,1}^{1/3}T_{\rm h,7}^{7/6}n_{\rm h,1}^{-1}$ pc, are formed in the shocked swept-up shell via TI.
A detailed description of assembly of clumps requires distribution function in the galactic potential (e.g., \citet{larson1969}), which goes beyond the scope of this Letter.
Here we focus on the total numbers of stars formed in the outflow and the corresponding star formation rate.

The cold clumps could be supported by turbulence with velocity dispersion $\sigstar=21.6\,\alpha_{\star} T_{\rm cl,1}^{-1/6}\\T_{\rm h,7}^{5/3} n_{\rm h,1}^{-1/2}\,\kms$ \citep{mckee2007}, where $\alpha_{\star}\sim1$.
In comparison, the adiabatic sound speed $\cs=0.3T_{\rm cl,1}^{1/2}\,\kms$.
We calculate the corresponding Jeans mass, $\mj$, which is the minimum mass to initiate gravitational collapse of the cloud, given by (see, e.g. \citet{mckee2007}):
\begin{equation}
\label{eq:eq14}
\mj=\left(\frac{3\pi^5}{32G^3\mpr}\right)^{1/2}\sigstar^3\,\ncl^{-1/2}\approx 3\times10^5\,T_{\rm cl,1}^{-1}\,T_{\rm h,7}^{9/2}\, n_{\rm h,1}^{-4}\,\msun\;.
\end{equation}
Comparing Eq.\ref{eq:eq9} and Eq.\ref{eq:eq14}, we have verified that $\mcl\sim10\,\mj$, indicating that TI-induced clouds collapse to form stars rapidly on a free-fall timescale much shorter than outflow's dynamical timescale $\tdyn$, $\tff\sim(G\rho)^{-1/2}\sim3\times10^4\,n_{\rm cl,7}^{-1/2}$ yrs, where $n_{\rm cl,7}=(\ncl/10^7\,\cmc)$.
The stars are deposited at the outflow speed nearly at rest along the outflow's path.
We assume that a fraction of the swept-up mass cools into form clumps and stars: $\dmstar\sim\fs\dms$.
The global star formation efficiency per dynamical time is often inferred to be around $\sim 1-10\%$ \citep{kennicutt1998, silk2013, somerville2015}, and so we adopt a total value of $\fs\sim10\%$ in the calculation.
In Fig.\ref{fig:fig1}, we find that the interception rate of swept-up mass by the outflowing shell for a MW mass halo, $\dms\sim10-100\,\msun\,\rm yr^{-1}$, corresponds to a star formation rate of $\sim 1-10\,\msun\,\rm yr^{-1}$, consistent with the recent observation of star formation within an AGN outflow \citep{maiolino2017}.
For halos of masses $\sim10^{11}\,\msun$, the estimated star formation rate drops to $\sim 0.1-1\,\msun\,\rm yr^{-1}$, while for halos of masses $\sim10^{13}\,\msun$, the expected star formation rate increases to $\sim 10-100\,\msun\,\rm yr^{-1}$.
The speed distribution of the newly born stars resembles the velocity profile of the outflowing shell, as shown in Fig.\ref{fig:fig1}.
Near the outer boundary of the galactic halo at $\sim 100$ kpc, the speeds of these stars exceed a few hundreds $\kms$, making them potential HVSs.
A fraction of these stars are unbound to the host galaxy.
%
%
\subsection{Statistics}
\label{subsec:stat}
\begin{figure}[h!]
\label{fig:fig3}
\centering
\includegraphics[angle=0,width=1\columnwidth]{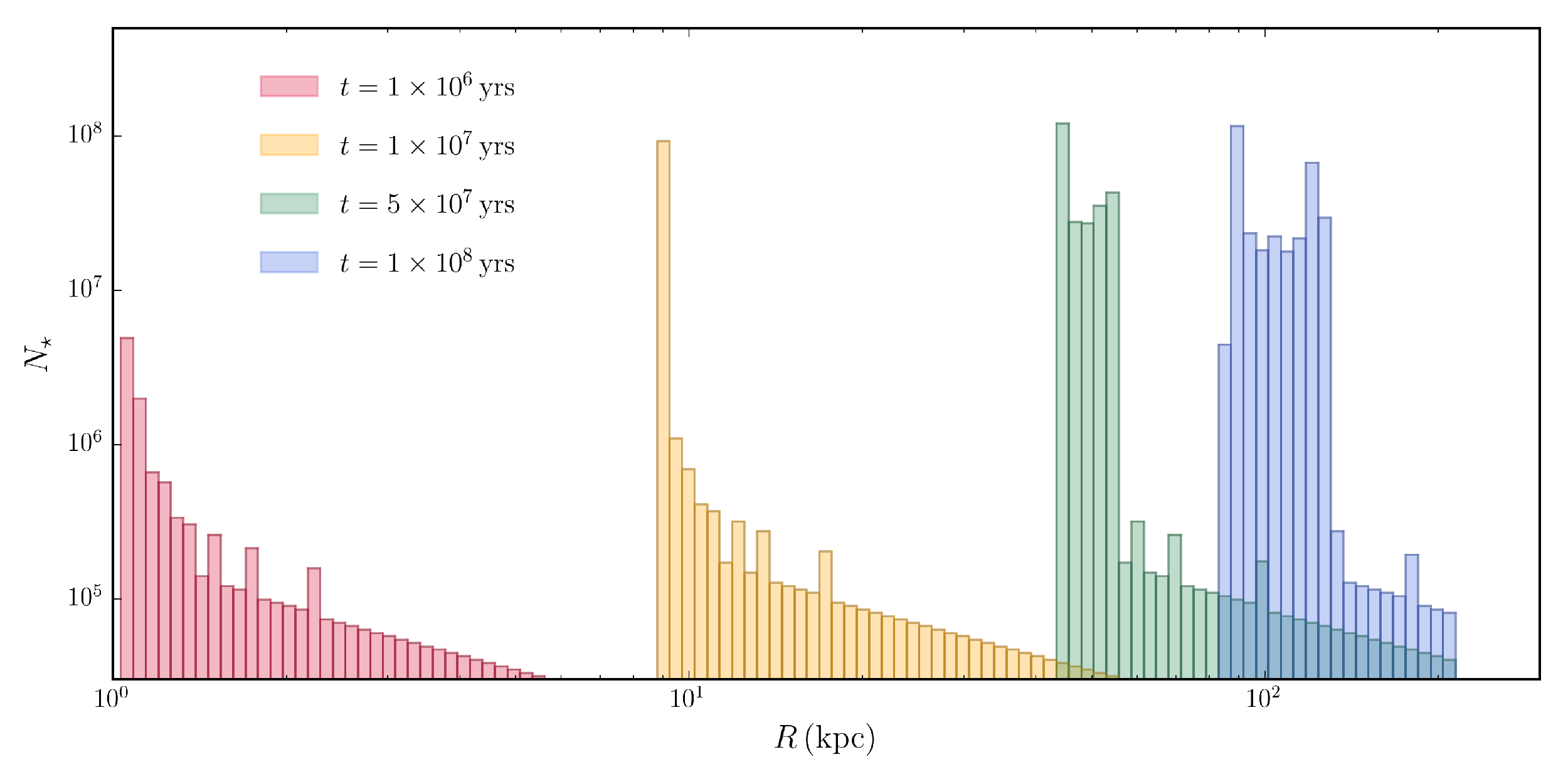}
\caption{
Cumulative number of stars in shells deposited along the outflow's path.
We divide the outflow into 150 logarithmically spaced shells and count the total number of stars found in each shell at a given time.
The red, orange, green and blue colors reflect the time elapsed since the outflow has launched.
We assume a star formation efficiency of $\fs\sim10\%$.
Stars are deposited with the speed of the outflow at the radial location and time of their birth.
Stars produced at early times could move ahead of stars produced later on in the outflowing shell.
}
\end{figure}

We divide the outflow's passage throughout the halo into a sets of shells of logarithmically equal width.
The number of stars produced per unit logarithmic radius in the outflow can be written as:
\begin{equation}
\frac{dN_{\star}}{d\ln\rs}=\frac{1}{\meanms}\frac{\dms\rs}{\vs}\;,
\end{equation}
where $\meanms$ is the average stellar mass derived from the Salpeter mass function, given by $\meanms=[(1-\beta)(1-s^{2-\beta})/(2-\beta)(1-s^{1-\beta})]\,\mmin$, where $s=\mmax/\mmin$, $\mmin=\msun$ and $\mmax=\infty$ are the minimum and maximum masses of stars, respectively.
For $\beta=2.35$, $\meanms\approx3.85\,\msun$.
We calculate the cumulative number of stars deposited in each shell at a given snapshot, shown in Fig.\ref{fig:fig3}.
The lifetime of a main-sequence star can be simply expressed as \citep{meurs1989}:
\begin{equation}
\taums=10^{\alpha}\left(\frac{M_{\star}}{\msun}\right)^{\beta}\;,
\end{equation}
where $\alpha=10.03$, $\beta=-4$ for $M_{\star}<1.5\msun$, $\alpha=9.86$, $\beta=-3$ for $1.5\msun\le M_{\star}<3.8\msun$, $\alpha=9.28$, $\beta=-2$ for $3.8\msun\le M_{\star}<12\msun$ and $\alpha=8.20$, $\beta=-1$ for $M_{\star}>12\,\msun$.
We have verified that stars of mass $\lesssim 2\msun$ will remain as main-sequence stars as the outflow reaches the edge of the halo.
Stars of mass $M_{\star}\gtrsim10\msun$ leave the main-sequence within the outflow's dynamical timescale, $\tdyn$.
The observed HVSs in the MW halo are massive B-type stars that are short-lived on main-sequence with lifetimes $\lesssim 10^8$ yrs \citep{brown2014}.
These stars fade at later times.
We estimate that $\sim 10^7-10^8$ HVSs are produced per MW galaxy during its entire lifetime.

During AGN periods, the predicted instantaneous HVS formation rate is $\sim 1-10\,\msun\,\rm yr^{-1}$, which is 4-5 orders of magnitude greater than the time-averaged rate of producing HVS via tidal breakup of binary stars \citep{hills1988} or three-body interaction between a star and a binary black hole \citep{yu2003, guillochon2015}.
Since the lifetime of bright AGN is of order a percent of the age of the Universe (e.g. \citet{martini2004}), the net production of HVS by AGN exceeds that from other mechanisms by several orders of magnitude, even when taking account of their short duty cycle.
%
%
\section{Summary \& Discussion}
\label{sec:summary}
In this Letter, we studied star formation in AGN outflows as a new mechanism for HVS production.
This possible channel for star formation is suggested by recent observations \citep{maiolino2017}.
We showed that the shocked ambient medium cools quickly and condenses to form cold clumps embedded in a hot tenuous gas via a thermal instability.
Stars are deposited along outflow's path at the local outflow speed.
We find that at a distance of $\sim 50-100$ kpc, stars are ejected with a speed of $\gtrsim 500\,\kms$ at a rate of $\sim 1-10\,\msun\,\rm yr^{-1}$, assuming a star formation efficiency of $\sim 10\%$.
Such a speed distribution is consistent with the HVS population in the MW halo \citep{brown2015}. 
During active periods of AGN outflows, the estimated HVS production rate is 4-5 orders greater than the rate predicted by other mechanisms.
We note that a more precise estimation requires a more realistic outflow geometry, which is beyond the scope of this paper.
However, a spherically symmetric outflow model adopted here produces hydrodynamic results consistent with observations of molecular outflows \citep{cicone2014,tombesi2015}.

The discovery of Fermi bubbles suggests that Sgr A* was recently active \citep{su2010}.
Our model predicts significant HVSs production from an outflow driven by such an AGN activity.
The observed B-type HVSs have lifetimes $\approx 10^8$ yrs, which suggests that Sgr A* could have been active $\sim 10^8$ yrs ago.
Stars formed during AGN episodes will be challenging to identify as they fade and travel to greater distances.
Recent IFU observations on local Seyfert galaxies with strong outflows revealed complicated gas structure and dynamics at $\sim$ kpc scales \citep{karouzos2016}.
Searches for HVSs in these galaxies could be promising provided that the local ionizing source of star formation dominates over AGNs, which can be justified by BPT diagram of emission line ratios \citep{baldwin1981}.

Stars born at early times during the outflow history travel faster than those formed at later times.
Thus, a large-scale double shell structure could appear, in which the outer shell consists of stars formed earlier while the inner shell contains stars formed later.
Outflows could also lead to the appearance of ring galaxies \citep{maccio2006} by clearing out halo gas \citep{zubovas2012} and producing a bright shell of stars at a large distance.
The shape of these configurations would reflect the three-dimensional geometry of the outflow.
Star formation rings are also predicted in cases where clouds are stationary before being struck by AGN winds, such as clumps within clouds struck by winds \citep{zubovas2014,  dugan2017}, clumps with high-velocity gas caused by the compression of the clouds from the outflow \citep{cresci2015}, and stars formed in giant molecular clouds within AGN winds \citep{tremblay2016}.
The resulting stellar distribution from those scenarios is different from the scenario we discuss where the resulting stellar population have the speeds of the outflow at birth.
%
%
\section*{Acknowledgements}
We thank Warren Brown and James Guillochon for insightful comments on the manuscript.
This work was supported in part by the Black Hole Initiative, which is funded by a grant from the John Templeton Foundation.
%
%

%
%
\end{document}